# Nanoporous Aluminum-Magnesium Alloy for UV enhanced spectroscopy

*Paolo Ponzellini[1], Giorgia Giovannini[1], Sandro Cattarin[2], Remo Proietti Zaccaria[1,3], Sergio Marras[1], Mirko Prato[1], Andrea Schirato[4], Francesco D'Amico[5], Eugenio Calandrini[1], Francesco De Angelis[1], Wei Yang[6], Hai-Jun Jin [6], Alessandro Alabastri[4], and Denis Garoli[1]\**

[1] Istituto Italiano di Tecnologia, via Morego 30, I-16163, Genova, Italy.
[2] ICMATE - CNR, Corso Stati Uniti 4, 35127 Padova, Italy.
[3] Cixi Institute of Biomedical Engineering, Ningbo Institute of Industrial Technology, Chinese Academy of Sciences, 1219 Zhongguan West Road, Ningbo 315201 P.R. China.
[4] Electrical and Computer Engineering, Rice University, 6100 Main Street MS-378, Houston, TX 77005
[5] Elettra Sincrotrone Trieste in Area Science Park, S.S. 14 Km 163,5 34012 Basovizza (TS) Italy
[6] Shenyang National Laboraory for Materials Science, Institute of Metal Research, Chinese Academy of Sciences, 72 Wenhua Road, Shenyang 110016 P.R. China

\* Corresponding author: Dr. Denis Garoli, denis.garoli@iit.it;

## Abstract

We report the first preparation of nanoporous Al-Mg alloy films by selective dissolution of Mg from a Mg-rich $Al_xMg_{1-x}$ alloy. We show how to tune the stoichiometry, the porosity and the oxide contents in the final film by modulating the starting ratio between Al and Mg and the dealloying procedure. The obtained porous metal can be exploited for enhanced UV spectroscopy. In this respect, we experimentally demonstrate its efficacy in enhancing fluorescence and surface Raman scattering for excitation wavelengths of 360 nm and 257 nm respectively. Finally, we numerically show the superior performance of the nanoporous Al-Mg alloy in the UV range when compared to equivalent porous gold structures. The large area to surface ratio provided by this material make it a promising platform for a wide range of applications in UV/deep-UV plasmonics.

## Introduction

During the last decade, Localized Surface Plasmon Resonances (LSPRs) have been explored extensively for their various technological applications such as surface-enhanced Raman spectroscopy (SERS), metal-enhanced fluorescence (MEF), plasmon enhanced light harvesting, and photocatalysis.[1–7] Plasmonic applications have been mainly based on noble metals (*e.g.* Ag and Au) because of their good chemical stability even though their application is limited to the visible/ NIR range.[4,8,9] However, the advantages of extending plasmonic enhancements down to UV and deep-UV (DUV) wavelengths are drawing interests on alternative materials.[10–12] For example, UV and DUV excitations can be uniquely exploited to extend Raman spectroscopy to biomolecules with vanishing Raman cross sections in the visible and NIR regions.[13–16] Beside Magnesium, Gallium, Indium, and Ruthenium, Aluminum (Al) has been suggested as a promising plasmonic material in the UV and DUV regions[17–23] because its large plasma frequency leads to a negative permittivity (real part) down to wavelengths of ≈100 nm.[24,25] Aluminum also exhibits strong enhanced local fields owing to its high electron density (3 valence electrons per atom compared to 1 valence electron per atom in metals such as Au or Ag) and its overall optical properties make it an excellent material for UV nanoantennas,[20,26,27] DUV SERS,[28–31] light emission enhancement of wide-bandgap semiconductors,[23] improvement of light harvesting in solar cells, and UV MEF.[17,32] Al nanostructures are generally designed with the help of electron beam lithography (EBL) and focused ion beam (FIB) lithography in order to obtain well-controlled designs.[20,26] However, since very small nanostructures/nanogaps (5-10 nm) are required to achieve plasmonic resonances in the DUV, and considering the long fabrication processes involved, these top-down techniques are not cost-effective and not recommended for large area fabrication (cm$^2$).[20,26] Several bottom-up approaches have been attempted in order to circumvent these difficulties, like nanoimprint lithography,[31] electrochemical anodization,[18] and chemical synthesis of aluminum nanocrystals.[33,34] Among the nanostructured/nanofabricated films, porous metals have recently attracted increasing interest due to their very high total exposed area. While several examples of porous metal applications exist in the Vis / Infrared spectral regions,[35,36] studies in the UV region have not been consistently reported so far. Here we show that nanoporous Al-Mg alloys (NPAM) can be prepared from an Al$_x$Mg$_{1-x}$ alloy by means of chemical dealloying procedures. We also confirm its performance in the UV measuring the MEF from Coumarin excited at 360 nm and the SERS signal from Adenine irradiated at 257 nm. We finally utilize actual cross-sectional SEM images of the fabricated NPAM films as the input geometry for electromagnetic calculations which highlight the effectiveness of NPAM in terms of electric field penetration and enhancement within the material cavities.

## Results and discussion

*Morphological and structural characterizations*

Chemical dealloying is a classical approach to prepare nanoporous metals with intriguing plasmonic properties. The most investigated material is probably nanoporous gold (NPG), [35–39] that is typically prepared by selective Ag dissolution from a $Ag_xAu_{1-x}$ alloy film deposited by means of co-sputtering from silver and gold targets. The same approach can be applied to aluminum, by depositing thin films composed of Al and a less noble metal and selectively leaching the latter by a judicious choice of the (mild) dealloying conditions. Unfortunately, the high reactivity of Al with atmospheric oxygen and moisture together with the tendency to passivation make this procedure extremely challenging. Magnesium is a good candidate partner for the preparation of a convenient alloy with Al, since: i) it can form a stable phase with Al (the γ-phase $Al_{12}Mg_{17}$)[40]; ii) it is less noble and easier to oxidize than Al, as required; and, notably, iii) it does not pose any significant toxicity hazard.

By tuning the respective sputtering powers acting on the magnesium and on the aluminum targets, we were able to tune the Al/Mg ratio in our $Al_xMg_{1-x}$ alloys (see table 1 and methods). The composition (obtained from EDS measurements) of the fabricated alloys is reported in Table 1, while the XRD analysis (see Fig. 2 and relative discussion) clarifies the nature of the sputter co-deposited alloys.

Our fabrication strategy considered the immersion of the deposited alloy samples in acid solution in order to dissolve Mg, either partially or completely, thus leaving nanometric or micrometric pores at its place. Several aqueous acid solutions were tested for the dealloying process, including: citric acid, ortophosphoric acid, sulfuric acid, as well as solutions of ammonium acetate and acetic acid in methanol. While all the tested solutions yielded nanoporous morphologies (see SI – Supporting Note#1), significant differences in the final oxide contents have been observed. Here we focus our analysis on the kind of samples prepared with 1M solution of $CH_3COOH$ in methanol, since such an etchant solution leads to the lowest level of residual oxide in the etched sample.

Table 1 reports the alloy compositions obtained before and after the chemical etching of representative samples named A, B, C. While sample A has not been treated in acid, the other two samples have been etched following the procedure reported in Methods. Their compositions are also reported (the recorded EDS spectra can be found in SI note#2). Importantly, oxygen contents between 10 and 28% have been achieved in all the etched samples. Noteworthy, considering that EDS may not be accurate enough to measure the oxygen content; the measured percentages have been verified by measuring the EDS spectra from a standard alumina sample ($Al_2O_3$) confirming the predicted aluminum and oxygen contents of about 40 and 60 % respectively. Interestingly, the non-complete etching of Mg from the alloy allowed for a reduction of the oxygen content. We do not expect that the presence of Mg would severely affect the plasmonic response of the NPAM films since Mg itself is known to be a good plasmonic material in the UV spectral region, although its high reactivity makes its use as standing-alone metal extremely challenging.[10]

**Table 1.** *Samples, initial composition parameter x, composition after dealloying as measured by EDS.*

| Sample | Sputtering power for Al–Mg (W) / Treatment | Pristine composition parameter x ($Al_xMg_{1-x}$)* | (EDS) Composition After etching (O – Al - Mg) |
|---|---|---|---|
| $A_0$ | 100-135 / no etching | 0.16 | ----- |
| $B_0$ | 100-85 / no etching | 0.23 | ----- |
| $B_1$ | 100-85 / 2 min etching | 0.23 | 12% - 23% - 65% |
| $B_2$ | 100-85 / 5 min etching | 0.23 | 28% - 30% - 42% |
| $C_0$ | 100-85 / annealing | 0.23 | ----- |
| $C_1$ | 100-85 / anneal., 5 min etching | 0.23 | 11% - 40% - 49% |

* For simplicity, Oxygen is not reported in this column. Oxygen levels in the pristine samples were always below 4%.

Fig. 1 reports the SEM micrographs of the samples. It is clearly shown that the nanoporous structure is achieved for all cases, even though the morphology strongly depends on the employed preparation conditions. Specifically, for sample $A_0$, Al and Mg have been co-sputtered for 50 minutes with applied powers of 100 W and 135 W, respectively. The obtained film resulted to be quite rough, a positive characteristic in terms of good UV plasmonic material.[17] For this reason we decided to measure its structural and optical behaviors without performing any etching procedure.

Vice versa, samples $B_0$, $B_1$, $B_2$; $C_0$ and $C_1$ were prepared by co-sputtering Al and Mg for 60 minutes with applied powers of 100 W and 85 W, respectively. For all these cases the obtained films were not as rough as sample $A_0$ therefore we proceeded with the etching step. Samples $B_1$ and $B_2$ were etched for 2 and 5 minutes, respectively, obtaining the morphologies depicted in Fig. 1($B_1$) and ($B_2$). In the 5 minutes etched sample the action of the acid is more evident. EDS shows that the amount of Mg diminished with the etching time, even though determining a higher level of oxidation (as shown even in the XPS analysis). Samples $C_0$ and $C_1$ (Fig. 1($C_1$)) were first annealed at 550 °C for 1 hour in the sputter coater, after the co-deposition, without breaking the vacuum. Then sample $C_1$ was obtained through a 5 minutes etching treatment. The SEM micrographs show how the etching process yielded much larger pores, in the micrometric scale. This morphology has been observed in all the samples (not shown) we annealed before the etching step. A possible interpretation for such a morphology is given in the discussion of the XRD data (below). The annealing step determines not only a different morphology in the etched sample, but also a different composition: samples $B_2$ and $C_1$ show quite different oxidation levels, despite being both etched for 5 minutes.

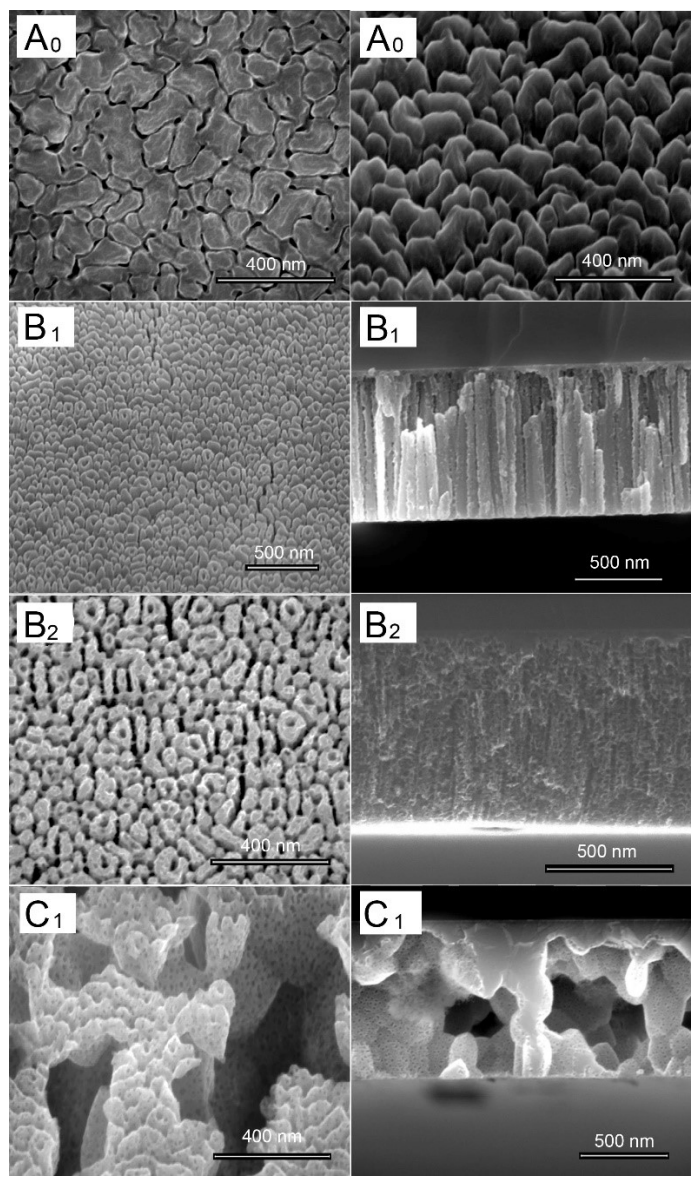

*Figure 1.* SEM micrographs of the prepared samples $A_0$, $B_1$, $B_2$ and $C_1$. The right column reports scans in tilted or cross-section view.

While through EDS we measured the bulk oxide content, we applied X-ray photoelectron spectroscopy (XPS) and X-ray diffraction (XRD) to better investigate the phase contents in the prepared films. These two compositional/structural analyses have been used to investigate the metallic nature of the obtained material. The XRD patterns obtained from the prepared samples are reported in Fig. 2. For those samples that underwent the etching treatment, spectra taken before and after etching are reported. Significant differences can be observed among the samples. In all cases, the structure of the not-etched layers ($A_0$, $B_0$, $C_0$) presents peaks of $Al_yMg_z$ stoichiometric phases, along with hexagonal phases that are prevalently composed of magnesium. The latter are substitutional solid solutions, where some smaller Aluminum atoms replace Magnesium atoms, with a resulting decrease of the lattice parameter. Since a smaller lattice parameter implies a higher diffraction angle, the peaks of these Mg-rich phases appear close to the peaks of pure Magnesium, but slightly shifted towards higher angles. It is interesting to notice the effect of the etching treatment on sample $B_0$. The XRD pattern of the pristine sample $B_0$ shows the $Al_{0.5}Mg_{1.5}$ phase and a Mg-rich phase (Fig. 2b). Upon etching, the XRD pattern shows: first a decrease of the Mg-rich phase ($B_1$), then its disappearance after prolonged treatment ($B_2$); no formation of a pure Al phase; decrease but persistence of the $Al_{0.5}Mg_{1.5}$ phase, with the main peak decreasing with the etching time in samples $B_1$ and $B_2$. As expected, the Mg-rich phase is more rapidly and effectively etched away. Let us now consider the effect of annealing. The XRD spectrum of sample $C_0$ shows (Fig.2c) that the annealing treatment causes an increase of crystallinity and the appearance of two distinct phases. The main peak appears intense, narrow and shifted to lower angles in comparison with Fig. 2b, very close to the Mg peak; this is evidence of a solid solution that is very-rich in Mg. Another peak, less intense but equally sharp, appears at larger angles and corresponds to the formation of a phase that is richer in Al, identified as the ordered intermetallic γ-phase ($Al_{12}Mg_{17}$).[40,41] As expected, the γ-phase shows good resistance to etching and its peak persists in sample $C_1$; conversely, the solid solution, very-rich in Mg, disappears completely upon etching. Within this framework, we can give a possible interpretation of the peculiar morphology of the annealed samples. The γ-phase and the Mg-rich solid solution tend probably to separate from one another, at the high temperature of the annealing treatment, forming large (micrometric) domains within the alloy layer. When the Mg-richer solid solution domains are etched, they leave large, almost micrometric pores in their place (Fig. 1($C_1$)), together with a secondary nanometric porosity. Noteworthy, in no one of the prepared samples any phase made only (or predominantly) of Al atoms could be detected. To conclude, the results from the XRD analyses suggest that the porous structures observed here can be due to a dealloying process that produce a selective phase etching.

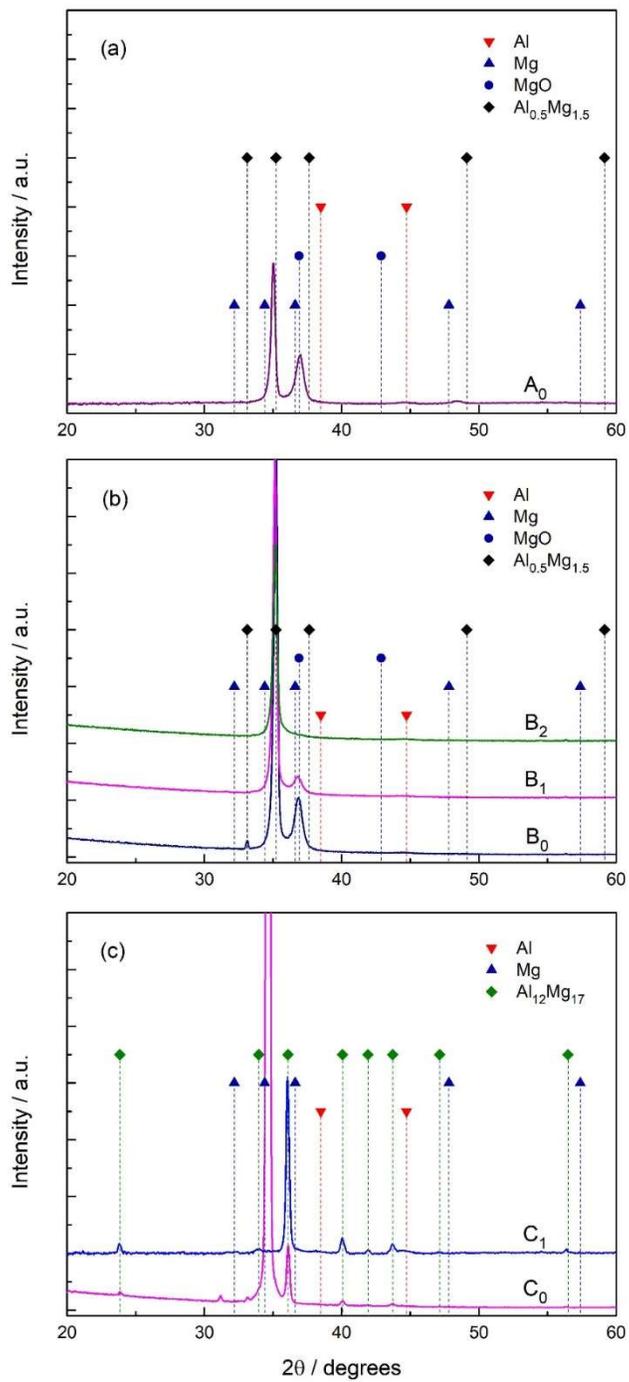

**Figure 2.** XRD analyses. (a) Sample $A_0$; (b) Pristine sample $B_0$ and the etched samples $B_1$ (etched for 2 minutes) and $B_2$ (etched for 5 minutes); (c) Annealed sample $C_0$ and the annealed, etched sample $C_1$ (etched for 5 minutes). (Only few peaks are present in the measured patterns because of the strong preferential orientation of the crystals within the Al-Mg layers.)

The strong reactivity of both Al and Mg, both in air and in the etching solutions, determines the oxidation of the surface of all the obtained samples. Even though both EDS and XRD investigations could demonstrate that the bulk composition is prevalently metallic, additional investigations have been performed in order to evaluate the surface film oxygen content. This analysis has been performed through XPS technique. Table 2 reports a summary of the obtained data (details of the measured spectrum are reported in SI note#2).

**Table 2.** *Samples, XPS analyses.*

| Sample | $Al_2O_3$ (at%) | Al suboxides (at%) | Metallic Al | MgO | Metallic Mg |
|---|---|---|---|---|---|
| $A_0$ | 5.2% | 0 | 5.8% | 71.0% | 18.1% |
| $B_1$ | 15.4% | 5.6% | 9.1% | 10.4% | 59.5% |
| $B_2$ | 40.7% | 4.5% | 10.6% | 4.8% | 39.4% |
| $C_1$ | 58.7% | 5.8% | 13.3% | 20.3% | 1.9% |

*For sample 135-100 pristine, the Al peak overlaps with the secondary peaks that are related to the high Mg content. The quantification may not be accurate, but some metallic Al is certainly present.

The results confirm what previously observed. A longer dealloying time implies higher surface oxidation, mainly in form of Al oxides. The pre-alloying annealing reduces the amount of metallic Mg in the final sample while slightly increasing the amount of Mg oxide. In all the samples, the metallic Al within the first few nm of the film is below 14% while higher amounts of metallic Mg are observed in samples $B_1$ and $B_2$. The main oxide, in all the etched samples, is $Al_2O_3$ while MgO seems to be dominant in the case of sample $A_0$

*Enhanced Spectroscopies – Plasmonic properties*

In the previous section we demonstrated the persistence of metallic phases in the etched samples. This finding, together with the obtained nanoporous morphology, suggests possible applications in enhanced spectroscopies. First of all, the reflectance of the prepared NPAM samples has been measured in the spectral range between 200 and 2000 nm. Due to the roughness of the samples, in order to collect the reflectance spectrum, a spectrophotometer equipped with an integrating sphere has been used. In this way the sum of the direct and diffuse reflectance could be collected. Supporting Note#3 (Fig. S13) reports a comparison between the reflectance of NPAM samples and Al and $Al_2O_3$ samples. Afterwards, the reflectance data have been used to evaluate, by means of a Drude-Lorentz fit, the dielectric constants of the samples (see SI note#3).

Finally, we investigated the potential plasmonic effects in terms of MEF and SERS.

In order to perform MEF experiments, all NPAM samples have been functionalized with a fluorophore according to the protocol reported in Methods. In particular, we chose coumarins as reporter dyes. They are well-known fluorophores widely exploited for their optical properties and for the development of *"off-on"* switchable fluorescent biosensors.[42,43] Specifically, 7-hydroxy-4-coumarin acetic acid was selected for the evaluation of the fluorescence enhancement (FE). This dye absorbs in the UV range (360 nm) and it has a suitable Stöke shift (100nm) which makes it easier to measure the fluorescence avoiding interferences related to the substrate itself. Furthermore, this dye already proved its applicability for the development of switchable sensors, therefore the possibility of enhancing its fluorescence signal could lead to intriguing improvements in terms of sensitivity and limit of detection (LOD) of fluorescent-based detection schemes.[44,45] A scheme of the different functionalization steps can be found in the SI (note#4, Figure S14). Similarly to Ray et al.,[17] we chose as the reference substrate a 10 nm rough Aluminum layer, covered with a 5 nm silica layer (methods). Such a substrate (named *"Al"* in the following) is reported to enhance the fluorescence signal.[17] In particular, we compared the FE obtained from substrates $A_0$, $B_1$, $B_2$ and $C_1$ with the value obtained from *Al*. The comparison was accomplished considering both the fluorescent signal measured for each substrate and the corresponding amount of dye effectively attached on the surface. In particular, we evaluated the enhancing efficiency by means of the FE factor defined as the ratio between the fluorescent signal and the calculated amount of coumarin on the surface. These values were finally normalized with respect to the one measured on sample *Al* (FE=1).

The FE value, the concentration of coumarin and the fluorescence signal determined for each substrate are reported in Figure 3. In order to compare the calculated concentration of coumarin attached on the surface and the correspondent measured fluorescence intensity, the relative values (normalized to the maximum measured value) for dye concentration (yellow bars) and fluorescence signal (blue bars) were considered. As noticeable, in spite of the highest concentration of coumarin for sample $B_1$ (relative value 100%; 147.50 µM), its measured fluorescence intensity is relatively low (relative value 24%; 4750 A.U.). This suggests that this sample has low enhancing properties (FE 1.4), a situation similar to that of the reference substrate, *Al* (FE 1). Even for *Al*, whereas the surface functionalization was efficient (71%; 105.03 µM), the fluorescent signal measured was 2410 A.U. (12% respect to sample $A_0$). With a FE value of 9.51, substrate $A_0$ is the one showing the best enhancing properties. Indeed, the highest fluorescent signal was measured for this sample (100%; 19788 A.U.) even though the amount of fluorophore attached on its surface is only 90.69 µM (the 61% of the max value, which was measured for sample $B_1$). Intermediate FE values were obtained for substrates $B_2$ and $C_1$, showing FE equal to 4.27 and 3.22, respectively. The measured/calculated values for each substrates are summarized in Table S1 of SI-note#4.

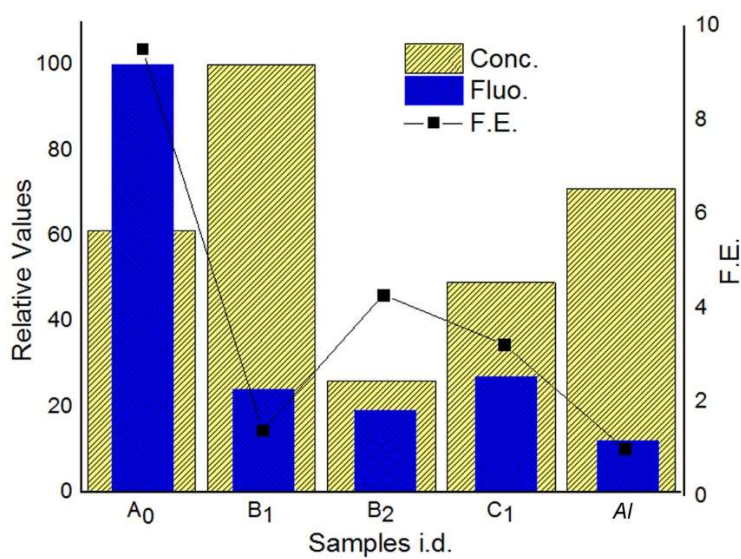

***Figure 3.** Relative values (normalized to the maximum measured value) of coumarin concentration (yellow bars) and fluorescence signal (blue bars). For substrates with good FE properties such as substrate $A_0$ (FE 9.51), low amount of dye (61%) lead to high fluorescence emission (100%). On the contrary, substrate $B_1$ showed low enhancing properties (FE 1.4) considering the low fluorescent intensity measured (24%) despite the high concentration of coumarin on the surface (100%).*

The good MEF property of substrate $A_0$ was confirmed by using $CsPbBr_3$ colloidal Nanocrystals (NCs), which can be excited at shorter wavelengths, *i.e.* 280 nm.[46] Even in this case, $A_0$ showed the better enhancing properties (FE 5.9) compared to *Al*. For more details see SI note#4, Figure S15.

The results obtained from MEF experiments are not surprising, as sample $A_0$ presents significant surface roughness (Fig.1) and, not being treated in acid, a low oxidation level. Nevertheless we would have expected more dye molecules to attach to the etched samples, which should have a much higher surface to volume ratio. Only $B_1$ shows a coumarin concentration that is higher than that of $A_0$ and Al. The lower dye concentrations measured for $B_2$ and $C_1$ might be related to the presence of a superficial oxide layer which makes the functionalization less efficient. Finally, even though the presence of the oxide layer is detrimental for MEF, we always obtained higher FE values with respect to the standard rough Al sample (*Al*). This is particularly more significant by recalling that rough Al is known to enhance the fluorescence, with respect to a quartz slide, by a factor of 2-9 depending on: the thickness (and hence roughness) of the Al layer,[17] the fluorophore-substrate distance,[47] the considered dye, and the possible presence of a second metal (Ag) in the substrate used for MEF.[48]

The capability of NPAM substrates to enhance the resonance Raman scattering in the UV was then evaluated. One of the most used test analyte for this kind of measurements is Adenine.[15,28,29,31] It is not only important in biology, as one of the nucleic acids, but it can be deposited as a thin film with a uniform thickness.[28] The samples were prepared by sublimating Adenine (99% purity, Sigma-Aldrich) on the NPAM samples under high vacuum ($\sim 8 \times 10^{-7}$ mbar). The adopted sublimation temperature was $\sim 180$

°C, which is far below the melting temperature of Adenine at atmospheric pressure (360 °C). The film thickness was monitored in situ by a calibrated quartz crystal microbalance. A uniform, thin (1 nm) Adenine film was formed on all the NPAM substrates prepared as described above, and on the same rough aluminum (*Al*) substrate used for the fluorescence experiment (methods). The experimental Raman spectra are presented in Figure 4.

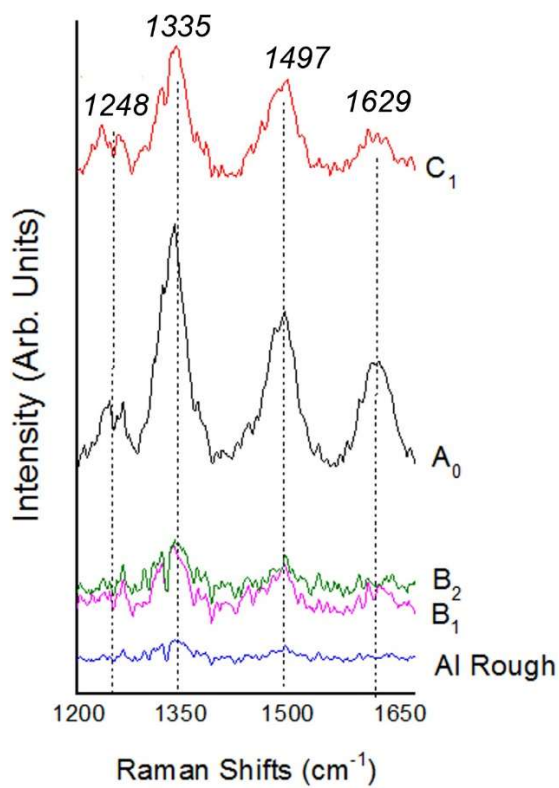

*Figure 4. SERS spectra of the adenine functionalized samples.*

The beam reaching the sample was approximately 2 µW of power, a much lower value than the mW scale typically used for lasers in conventional Raman spectroscopy setups. By this choice of numbers it was possible to measure at 257 nm exciting radiation and still to avoid photodamage phenomena. However, a drawback related to this approach needs to be stressed, namely the obvious low number of photons that can be collected and the consequent high noise level in the signal (see Fig. 4).

In the spectra, especially strong peaks were observed at 1248, 1335, 1497, and 1629 cm$^{-1}$ all attributable to Raman scattering of adenine vibrations.[28] All four NPAM substrates showed higher intensities of these bands compared to the rough Al substrate. The largest enhancements were observed with the substrates $A_0$ and $C_1$. For quantitative evaluation of the enhancement it was analyzed the intense and well-isolated band at 1497 cm$^{-1}$, attributable to the combination tone of $C_4N_9$ stretching and $C_8H$ bending vibrations[49]. Table 3 reports the SERS enhancement factor (SERS-EF) comparison between the different tested samples, where $A_0$ shows the largest achieved enhancement (about 10). Importantly, this result agrees with the MEF findings shown in Figure 3. Similarly, among the porous samples, $C_1$ shows the highest SERS-EF. Noteworthy, even though the reference rough Aluminum substrate is known to provide enhanced Raman signals in the UV region,[50] its SERS-EF value results to be much lower than NPAM substrates. This might be due to the different morphology between the Al film and NPAM where the latter are characterize by "hot-spots". However, even though it appears as a possible explanation more investigation is needed in this respect.

**Table 3.** *Samples, SERS Enhancement calculation.*

| Sample | Peak Area | SERS E.F. |
|---|---|---|
| *Al* rough | 700 | 1 |
| $A_0$ | 6600 | 9.5 |
| $B_1$ | 2000 | 2.8 |
| $B_2$ | 1000 | 1.4 |
| $C_1$ | 4000 | 5.6 |

*Numerical simulations*

In order to gain insights regarding the microscopic interaction between UV radiation and the fabricated material, we have combined the actual nanoporous film geometry of sample $C_1$ with 2D electromagnetic calculations to investigate the electromagnetic field localization within the NPAM film. The SEM image of a vertical cross section of a fabricated NPAM film (Fig. 5a) has been imported as weighted map (Fig.

5b) to obtain a realistic description of the optical response of the film (see Methods and SI-note#5 for further details). In Fig. 5c we show the electric field distribution in the NPAM film at λ = 260 nm and with the polarization parallel to the plane. The field penetrates for a few tens of nanometers within the nanoporous structure, generating hot spots near the edges of the Al-Mg alloy, within the pores. For comparison, in Fig. 5d, we show, on the same scale and for the same geometry, the electric field in case of NPG. While the profiles for both NPAM and NPG are similar, the NPAM film clearly exhibits larger peak values, confirming its field enhancing properties in the UV region. The comparison for additional excitation wavelengths can be found in the SI-note#5.

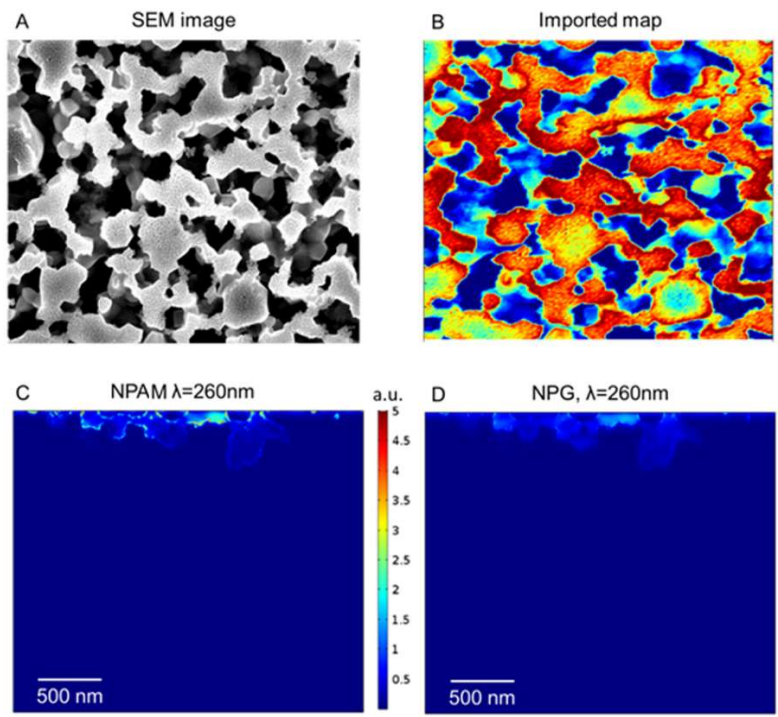

*Figure 5.* a) SEM image of vertical cross section of prepared nanoporous Al-Mg film. b) Imported map of the cross section weighted on the contrast of SEM image. c,d) e.m. calculations of field confinement (same scale) at a wavelength of 260 nm using the optical properties of NPAM and NPG films respectively.

**Conclusions**

In conclusion, here we have reported about the procedures to prepare NPAM films from an alloy of Al and Mg. Co-sputtering from Al and Mg allows to deposit films with different concentration ratios between Al and Mg. For all cases, the dealloying processes lead to a residual oxide layer. This layer is found to be stable in time and always covering a porous structure of metallic Al-Mg alloy. Optical spectroscopies have been used to evaluate the optical properties of the prepared NPAM films and to verify their potential applications as UV plasmonic materials. The results, in terms of bulk composition of the prepared NPAM samples, suggest that the final oxygen content, primary detrimental factor in the application of NPAMs as plasmonic materials, depends both on the starting concentration of Mg with respect to Al and on the residual amount of Mg in the etched alloy. During our investigation we observed that, while it is possible to remove completely the Mg contents from the starting alloy by means of strong / long lasting acid etching, this always leads to high oxygen contents pushing the final sample composition close to that of $Al_2O_3$. This is not however the case if the dealloying process is optimized in order to create a nanoporous structure without completely removing Mg, an approach that seems to effectively lower the oxidation level within the bulk porous structure. MEF and SERS experiments demonstrated that the prepared substrates operate much better than standard rough Al films. Not surprisingly, a highly rough Al-Mg film can indeed give a higher enhancement due to lower surface oxidation. Finally, the present study should be considered as the first step towards the identification of an optimized fabrication procedure capable of further reducing the final oxygen contents in NPAM samples which, in turn, would lead to a powerful platform for UV enhanced spectroscopy.

***Materials:*** 7-hydroxil-4-coumarin acetic acid 97%, N-(3-Dimethylaminopropyl)-N′-ethylcarbodiimide hydrochloride BioXtra (EDC hydrochloride), N-Hydroxysuccinimide 98% (NHS), (3-Aminopropyl)triethoxysilane 99% (APTES), acetic acid glacial 99.85%, Methanol anhydrous, 99.8%,

Dimethyl sulfoxide anhydrous, ≥99.9% (DMSO). All mentioned chemicals were purchased from Sigma-Aldrich. Acetone, anhydrous (max. 0.01% $H_2O$) ≥99.8% was purchased from VWR chemicals.

*Methods:*

*Sputter deposition:* silicon substrates have been sonicated in acetone and isopropanol to obtain a clean surface. A 10 nm Chromium layer and an Aluminium layer of the same thickness have been deposited on the silicon substrate, as the linker between the substrate and the subsequently deposited Al-Mg alloy. Al and Mg have been then sputter co-deposited by applying different biases to the Aluminium and Magnesium targets, as described in the text. All these processes have been performed in Argon atmosphere, at 5 µBar, in a Kenosistec KS500 Confocal sputter coater. The chamber was set to high vacuum (P<$10^{-6}$ mBar) before the deposition of every layer. Chromium and Magnesium were deposited in DC, while Aluminum in RF.

*Rough Al evaporation:* Aluminum and Silica have been evaporated within a Lesker E-beam evaporator. The chamber was set to high vacuum (P<$10^{-6}$ mBar) before the deposition. The 10 nm Aluminum layer has been deposited at 0,2 Å/s, the 5 nm silica layer at 1 Å/s. The silica layer was deposited just after the Aluminum layer, without losing the vacuum in the evaporator chamber.

*Etching by acid treatment:* sample $B_1$ was dipped in a 1 M methanol solution of acetic acid for 2 minutes while substrate $B_2$ and $C_1$ for 5 minutes. The samples were then washed with methanol, isopropanol and pentane. Immediately after the acidic treatment, the samples were brought inside the glove box.

*X-ray diffraction (XRD)* patterns were recorded on a PANalytical Empyrean X-ray diffractometer equipped with a 1.8kW CuKα ceramic X-ray tube, PIXcel3D 2x2 area detector and operating at 45 kV and 40 mA. The diffraction patterns were collected in air at room temperature using Parallel-Beam (PB) geometry and symmetric reflection mode. XRD data analysis was carried out using HighScore 4.18 software from PANalytical.

*X-ray photoelectron spectroscopy (XPS)* analysis has been carried out with a Kratos Axis UltraDLD spectrometer (Kratos Analytical Ltd., UK) using a monochromatic Al Kα source (hν = 1486.6 eV) operated at 20 mA and 15 kV. The analyses have been carried out on 300 × 700 μm area. High-resolution spectra have been collected at pass-energy of 10 eV and energy step of 0.1 eV. Spectra have been analyzed with CasaXPS software (Casa Software, Ltd., version 2.3.17)

*Samples functionalization with APTES:* All substrates ($A_0$, $B_1$, $B_2$, $C_1$ and *Al*) were shacked overnight, at RT, in 4% solution of APTES in acetone inside glove box. All substrates were then washed in acetone and dried inside the glove box.

*Dye activation and attachment on the substrate*: 7-hydroxyl-4-coumarin acetic acid (0.01 mmol) was solubilized in DMSO (1 mL) reaching 10 mM as final concentration. EDC (0.02 mmol) was added and stirred at room temperature for 15 minutes. Subsequently, NHS (0.03 mmol) were added and the reaction was stirred for further 45 minutes. After 1 hours, the mixture was diluted in acetone reaching a final solution of 1 mM of dye. 1 mL of this solution was added to each substrate in a 3 mL glass vial and 1 mL was added in an empty vial as reference for the dye quantification. The reaction was shacked at R.T. in glove box overnight. The substrates were carefully withdraw and washed with acetone. The remaining solution of coumarin was kept for dye quantification: the absorbance at 360 nm was used to determine the amount of coumarin present in each solution using as reference 1 mM solution of activated coumarin. The value founded were used to calculate the amount of coumarin covalently bound on the substrate's surface.

*Fluorescent measurement*: after washing all substrates were placed in a 24-well plate, a drop of PBS was added on the surface in order to make a measurement under wet condition, and the fluorescent spectrum was recorded with a 360 nm excitation wavelength ($\lambda_{ex}$). Fluorescent signal was measured selecting specific section of each well ($\lambda_{ex}$:$\lambda_{em}$ 360:460 nm) in order to better evaluate the efficiency of the functionalization. Tecan infite M200 was used for recording the fluorescent spectrum and for the fluorescent measurement of the substrates.

*CsPbBr₃ colloidal Nanocrystals (NCs) deposition and measurement:* In glove box (argon atmosphere, $H_2O$ < 0.1 ppm; $O_2$ < 0.1 ppm, RT.) 2.5 μL of NCs solution (6.5 x $10^{14}$ NCs/mL in toluene) were dropped on substrate $A_0$, silicon and *Al* ensuring that all the suspension remain on the substrate to guarantee that the same amount of NCs are deposited on all substrates. After solvent evaporation, the fluorescence signal was measured for each substrates using Tecan infinite M200, using 280:510 nm as $\lambda_{ex}$:$\lambda_{em}$ respectively.

*RAMAN*: UV Resonant Raman (UVRR) measurements have been carried out at Elettra synchrotron radiation facility. A complete description of the experimental apparatus can be found elsewhere.[51] The fine wavelength tunability of the synchrotron based excitation source allowed to set the UVRR excitation source at 257 nm. Such wavelength corresponds to the maximum of the UV-Vis adenine absorption spectrum. This way it is possible to fit the best resonant condition within the adenine molecules. The Raman scattering signal was collected with a backscattering configuration. The beam reaching the sample was approximately 2 μW of power. A Czery-Turner spectrometer with focal length of 750 mm, coupled with an holographic reflection grating of 1800 g/mm and with a Peltier-cooled back-thinned CCD was employed to get the Raman signal. Spectral resolution was set to 25 $cm^{-1}$ in order to get a sufficient count rate. Raman frequencies were calibrated to ± 1 $cm^{-1}$ of accuracy by using cyclohexane spectra.[52]

*FEM simulations*: Numerical simulations have been performed via a finite element method model, developed using COMSOL Multiphysics 5.3a. Maxwell's equations have been solved on a two-dimensional geometry modelling the sample to determine the field enhancement within the structure.
In particular, since the simulations aimed at investigating the field localised on nanopores and concave shapes of the surface, a faithful modelling of the metal layer is crucial. Therefore, as any numerical structure would not thoroughly reproduce the experimental sample, SEM images of the porous metal section have been directly employed. The measured profiles have been imported as images on COMSOL and used as permittivity maps associating each point $(x, y)$ to a value between 0 and 1 according to the SEM image brightness. As a result, $map(x, y) = 1$ represents pure metal, $map(x, y) = 0$ background only. On the other hand, concerning nanoporous irregular metal surface, such a layer has been associated to a weighted permittivity defined as follows:

$$\varepsilon = \begin{cases} \varepsilon_{metal} * map(x, y) \text{ if } map(x, y) > ths \\ \varepsilon_{background} \text{ if } map(x, y) < ths \end{cases}$$

where a threshold value $ths$ has been defined to distinguish background medium from metal and model the surface boundaries. In the simulations surrounding environment is air, with $\varepsilon_{air} = 1$, while substrate is the investigated metal – either Al-Mg or Au. The permittivity of the alloy used for the simulations has been obtained from ellipsometric measurements on the as-deposited alloy (B$_0$),[24] whereas Au has been optically described according to reference[53]. In addition, $\varepsilon$ has been weighted by the value of the map itself in order to smoothen the edges of media.

The threshold has been tuned to reproduce the experimental sample as much as possible. Indeed, $ths$ directly affects the permittivity governing the optical properties of the porous layer, determining the metal boundaries according to the SEM map values. In particular, the selected value for the field maps presented in Fig. 5 is $ths = 0.4$. Importantly, the threshold value is common for simulations of both Al-Mg and Au, which guarantees a no-biased comparison between performances of the two metals. Nevertheless, tests at different values of $ths$, varying between 0.2 and 0.9, have been conducted for the both Al-Mg and Au.

Thus, after having appropriately set the domain optical properties within the model geometry, Maxwell equations have been solved to determine the field spatial distribution. The incoming electric field is simulated by means of a port analysis, described as a plane wave at a given wavelength. Its amplitude is unitary, and its propagation direction is orthogonal to the surface while polarization is parallel to the sample. Continuous periodic conditions are imposed at the two sides of the system, and perfectly matched layers have been defined at the top and bottom of the structure, to prevent spurious enhancement due to unphysical reflections at the boundaries of the domain. Concerning mesh, the region corresponding to the interface between metal and environment required a particularly fine mesh to resolve the electric field within the nanopores. Two-dimensional triangular elements have been employed, with $2.5\ nm$ and $10\ nm$ as minimum and maximum size respectively.

Furthermore, in order to investigate and compare Al-Mg plasmonic behaviour with Au response, the same SEM profile has been associated to the two different material permittivity. In addition, data of the field enhancement from nanoporous layers have been analysed in a fixed range only, for both metals. Indeed, by saturating the ranges of the resulting enhancement, one can straightforwardly compare the computed values and notice to what extent pores in Al-Mg can localise the field more efficiently than Au in the UV spectral range.

## AUTHOR INFORMATION

* Corresponding author: Dr. Denis Garoli, denis.garoli@iit.it;

## AUTHOR CONTRIBUTION

PP, GG, fabricated and characterized the structures; SC conceived the dealloying method; SM and MP performed the structural analyses; FD helped in the optical characterization; RPZ, AS and AA performed the e.m. simulations; EC helped in the optical spectroscopy analyses; WY, JH, RPZ and FDA helped in manuscript preparation. PP, GG and SC equally contributed to this work. DG supervised the work.

## ACKNOWLEDGMENTS

The research leading to these results has received funding from the European Research Council under the Horizon 2020 Program, FET-Open: PROSEQO, Grant Agreement n. [687089]. The authors want to thank Liberato Manna, Dmitry Baranov and Guilherme Almeida for their support in the colloidal nanocrystal preparation.